\documentclass[aps,prl,twocolumn,superscriptaddress,amsmath,amsfonts,showpacs]{revtex4} 

\usepackage{graphicx}
\usepackage{bm} 
\usepackage{amssymb,amsmath}

\begin{document}


\title{
Majorana Fermions in Nodal Superconductors: Gapless Topological Phase
}


\author{Masatoshi Sato}
\affiliation{The Institute for Solid State Physics, The University of
Tokyo, Chiba 277-8581, Japan}
\author{Satoshi Fujimoto}
\affiliation{Department of Physics, Kyoto University, Kyoto 606-8502, Japan}




\date{\today}

\begin{abstract}
We demonstrate that Majorana fermions exist in edges of systems and in a vortex core
even for superconductors with nodal excitations such as the $d$-wave pairing state
under a particular but realistic condition in the case with 
an anti-symmetric spin-orbit interaction and a nonzero magnetic field below the upper critical field.
We clarify that the Majorana fermion state is topologically protected 
in spite of the presence of bulk gapless nodal excitations, because of the existence of
a nontrivial topological number.
Our finding drastically enlarges target systems where we can explore the Majorana fermion state.
\end{abstract}



\maketitle



{\it Introduction ---}
Majorana fermions realized as vortex core bound states of superconducting condensates have been attracting
considerable interest in connection with the application to quantum computation\cite{freedman,das,kitaev2}.
Such vortices obey the non-Abelian statistics\cite{moore,nayak,fradkin,read,ivanov,stone1}, and because of this distinct feature,
they are utilized as decoherence-free qubits.
The realization of Majorana bound states has been discussed for the quantum Hall effect systems\cite{moore,nayak,fradkin,read},
$p+ip$ superconductors\cite{read,ivanov,stone1,machida,cooper}, superconductor-topological-insulator interfaces\cite{FK08,LTYSN10}, and $s$-wave Rashba superconductors\cite{STF09,SAU,ALICEA}.
The origin of Majorana fermions acting as non-Abelian anyons is intimately related to the existence of the non-Abelian topological order, which yields the fractionalization of quasiparticles\cite{wen2,lee}.
Generally, topological order is characterized by a nontrivial topological number associated with
the global structure of the Hilbert space, and hence, the existence of a nonzero energy gap which separates
the topological ground state and non-topological excited states
stabilizes the topological order.

In this paper, we propose an example of systems realizing Majorana fermions, which is unusual in the above-mentioned sense of topological stability, but ubiquitous in real materials:
Majorana fermion states realized in superconductors with nodal excitations such as the $d$-wave pairing state. 
More precisely, Majorana fermions exist in edges of systems and in a vortex core
for nodal superconductors, provided that there are both an anti-symmetric spin-orbit interaction and a nonzero magnetic field below the upper critical field $H_{c2}$,
and that when the magnetic field is switched off, the Fermi level is located
close to odd numbers of time-reversal invariant $\bm{k}$-points 
in the Brillouin zone at which the superconducting gap vanishes because of symmetry requirement.
This proposal implies that the non-Abelian topological order coexists with gapless excitations in our system.
One may wonder how the topological stability is ensured in the presence of
non-topological gapless excitations.
In fact, the Chern number is not well-defined in our nodal system. 
Nevertheless, we clarify that the Majorana fermion state is, to some extent, stable against interactions
with nodal excitations and with impurities, because of the existence of a topological number
which is well-defined even for gapless superconductors.
Note that many classes of noncentrosymmetric (NCS) superconductors such as CePt$_3$Si, CeRhSi$_3$, CeIrSi$_3$, Li$_2$Pt$_3$B, 
are known to possess superconducting gap-nodes\cite{CePtSi2,CeRhSi,CeIrSi,mukuda,zheng}.
In these systems, some of time-reversal invariant $\bm{k}$-points reside 
close to the Fermi level\cite{harima}. Our finding indicates that 
if the total number of these $\bm{k}$-points is odd, and
the superconducting gap vanishes (or, at least, becomes sufficiently small) at these points,
stable Majorana fermion modes appear under applied magnetic fields.
We expect that such Majorana fermion states may be realized in large classes
of NCS superconductor with gap-nodes.

{\it Majorana fermions in edges and in a vortex core ---}
To be concrete, we consider a two-dimensional $d$-wave superconductor with the Rashba spin-orbit (SO)interaction, 
though the following argument is basically applicable to any NCS nodal superconductors.
The Hamiltonian is
given by
${\cal H}=\frac{1}{2}\sum_{\bm k}
\psi_{\bm k}^{\dagger}
{\cal H}({\bm k})
\psi_{\bm k}$,
with 
\begin{eqnarray}
{\cal H}({\bm k})=
\left(
\begin{array}{cc}
\epsilon_{\bm k}-h\sigma_z+{\bm g}_{\bm k}\cdot{\bm \sigma}
& i\Delta_{\bm k} \sigma_y\\
-i\Delta_{\bm k}\sigma_y
& -\epsilon_{\bm k}+h\sigma_z+{\bm g}_{\bm k}\cdot {\bm \sigma}^{*}
\end{array}
\right),
\label{eq:hamiltonian}
\end{eqnarray}
where $\psi^{\dagger}_{\bm k}=(c^{\dagger}_{\bm{k}\uparrow},c^{\dagger}_{\bm{k}\downarrow},c_{-\bm{k}\uparrow},c_{-\bm{k}\downarrow})$, $\epsilon_{\bm k}=-2t(\cos k_x +\cos k_y)-\mu$,
${\bm g}_{\bm k}=2\lambda(\sin k_y, -\sin k_x,0)$,  
$\bm{k}=(k_x,k_y)$,
and $\bm \sigma=(\sigma_x,\sigma_y,\sigma_z)$ 
the Pauli matrices. $h=\mu_{\rm B}H_z$ is a Zeeman magnetic field.
The gap function is the $d_{x^2-y^2}$-wave type, $\Delta_{\bm k}=\Delta_0(\cos k_x-\cos k_y)$, 
or the $d_{xy}$-wave type, $\Delta_{\bm k}=\Delta_0\sin k_x\sin k_y$.
We neglect the orbital effect of the magnetic field for a while
since it does not change our results qualitatively,
as long as $H_z<H_{c2}$. 

\begin{figure}[h]
\begin{center}
\includegraphics[width=7cm]{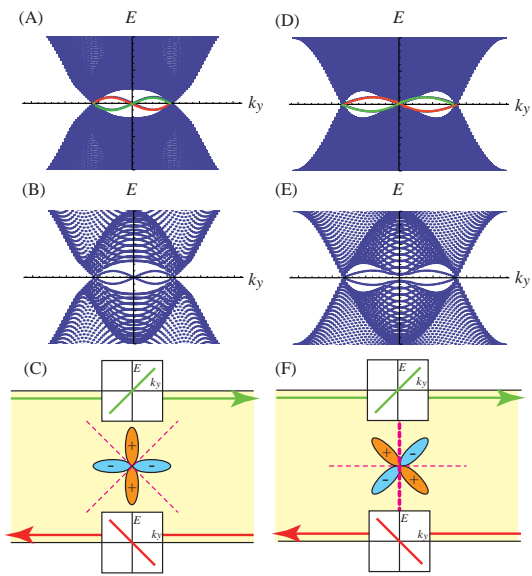}
\caption{Energy spectra for systems with open boundaries for the
$x$-direction and the periodic boundary condition for the
$y$-direction. $\mu=-4t$, $\lambda=0.5t$, $\Delta_0=t$ and $h=2t$. 
(A) (B) $d_{x^2-y^2}$ wave pairing. 
(D) (E) $d_{xy}$ wave pairing.
The distance between two edges is $L=90$ [(A) (D)] and $L=30$ [(B) (E)]. In (A) and (D),
chiral gapless edge modes at $x=0$ and $x=L$ are depicted, respectively, in green and red curves.
(C) (F)  Chiral edge modes counter-propagating on two opposite edges for $d_{x^2-y^2}$ wave pairing (C) and $d_{xy}$ wave pairing (F).
}
\label{fig:edgestate}
\end{center}
\end{figure}

We, first, demonstrate that there is a gapless chiral Majorana fermion mode on the edge of the system.
For this purpose, we numerically calculate the energy  spectrum of the system 
with the open boundary condition imposed for the $x$-axis, and the periodic boundary condition for the
the $y$-axis.
The results are shown in FIG.\ref{fig:edgestate}.
In the case of the $d_{x^2-y^2}$-wave pairing, when the condition $-4t-\mu<h<-\mu$ is satisfied,
a gapless edge mode appears for $k_y\sim 0$. (FIG.\ref{fig:edgestate}(A))
Note that this edge mode is isolated from
continuum of gapless excitations from the gap nodes with the finite Fermi momentum.
Because of the particle-hole symmetry of the Hamiltonian,
the existence of one zero energy mode implies that
the zero energy edge mode for $k_y\sim 0$ is a Majorana fermion.
The above condition implies that when the Fermi level crosses 
$\bm k$-points close to time-reversal invariant points in the Brillouin zone
in the absence of the magnetic field, say, the $\Gamma$ point; i.e. $\mu=-4t$,
the Majorana edge mode appears for any nonzero $H_z$ below $H_{c2}$.
This property makes a sharp contrast to the $s$-wave pairing state considered in refs.\cite{STF09,SAU,ALICEA}, 
for which
a large magnetic field satisfying $h>\Delta$ is required to realize the topological order and Majorana fermions.
%
This is because that 
the gap function of the $d$-wave pairing vanishes at $\bm{k}\sim 0$, fulfilling
the condition $h>\Delta_{\bm{k}\sim 0}\sim 0$.
As a result, the realization of Majorana fermions in the $d$-wave pairing state is much more feasible than
that in the $s$-wave pairing state, which may be seriously affected by the orbital depairing effect
due to the large magnetic field $h>\Delta$.

In the case of the $d_{xy}$-wave pairing, the zero energy edge mode at $k_y\sim 0$ merges into
nodal excitations, and thus, it is difficult to identify the Majorana
mode from the numerical result (FIG.\ref{fig:edgestate}(D)).
However, we deduce that there is still a Majorana mode at $k_y\sim 0$ also for the $d_{xy}$-wave pairing,
because of the topological argument which will be presented later, as
long as $L$ is sufficiently large. (also see the argument on the stability below.)

The existence of the chiral Majorana fermion edge mode implies
that there is a Majorana fermion mode in a vortex core of the superconducting condensate when
the vorticity $n$ is odd.
In the case of the $d$-wave pairing, the analysis of
the vortex core state is cumbersome, 
in contrast to the $s$-wave pairing or the $p+ip$-wave pairing states,
since the gap function of the $d$-wave state is not an eigen state of the orbital angular momentum,
and, moreover, there is no truly localized bound state in a vortex core because of interactions with delocalized
nodal excitations\cite{franz,franz2}.
However, 
as in the case of the $s$-wave Rashba superconductor\cite{STF09,STF10},
we can construct the Majorana zero energy mode from quasiparticles with $\bm{k}\sim 0$
at least in some parameter regions. 
The eigen function for this zero energy state is
$\phi^{T}=(u_{\uparrow},u_{\downarrow},u_{\uparrow}^{*},u_{\downarrow}^{*})$ with 
$u_{\uparrow}=ie^{i\frac{n-1}{2}\theta}f(r)$, 
$u_{\downarrow}=-ie^{i\frac{n+1}{2}\theta}f(r)$
and $f(r)=\sqrt{\frac{h}{\pi\lambda r}}e^{-\frac{h}{2\lambda}r}$ for large $r$.
Here $n$ is odd.
In the $d$-wave pairing state, in addition to this zero energy mode, 
there are four gapless extended states outside of the vortex core which stem from the four gap-nodes\cite{volovik,melnikov}.
Since the total number of the zero energy mode is odd, one Majorana mode survives.
Thus, we have a zero energy Majorana fermion mode in the vortex core.

{\it Stability of Majorana fermions ---}
The next important question is whether Majorana fermions found above
are stable or not against weak perturbations such as impurities
even in the presence of gapless nodal excitations.
As will be shown later, there is indeed a topological protection mechanism
in spite of the existence of bulk gapless excitations.
Before discussing the topological mechanism, however, we here present a heuristic argument on this issue
to grasp an intuitive physical picture.
Generally, impurity scattering affects the superconducting state with gap-nodes.
We consider only the case that the density of impurities is sufficiently small so that
the superconducting gap is not much reduced.
We, first, consider the case of the $d_{x^2-y^2}$-wave pairing. 
In a semi-infinite system with an open boundary, 
there is only one chiral Majorana edge mode, while there are four gapless modes
which stem from four nodes of the $d_{x^2-y^2}$-wave superconducting gap.
To generate an energy gap in the Majorana spectrum, we need even numbers of Majorana modes
which are paired into complex fermions.
Thus, for this geometry, the chiral Majorana edge mode is stable against interactions with
nodal excitations, and also against impurity scattering.
However, this argument is not applicable to the case with
two open boundaries at the opposite sides of the system.
In this case, two counter-propagating chiral Majorana modes reside
in the two opposite edges, as depicted in FIGs.\ref{fig:edgestate}(C).
Interactions between bulk gapless nodal excitations and two chiral Majorana modes
may give rise to long-range tunneling between two Majorana modes.
We note that such long-range tunneling via nodal excitations does not occur in a clean system,
because of the mismatch of the Fermi momenta of nodal excitations
$k_F\neq 0$ and that of the chiral Majorana modes with which $k_F\sim 0$
for the $d_{x^2-y^2}$-wave pairing.
When there are impurity potentials, 
a Majorana mode and nodal excitations on an edge can be hybridized via impurity scattering,
leading to
the long-range tunneling of the Majorana fermions in two opposite edges:
$\mathcal{H}_{\rm tun}=t(\bm{r}_0-\bm{r}'_0)i\gamma(0,y_0)\gamma(L, y'_0)$,
where $\gamma(x,y)$ is a Majorana fermion operator,
 $\bm{r}_0=(0,y_0)$ and $\bm{r}_0'=(L,y'_0)$ are the positions of impurities 
 and the tunneling amplitude $t(\bm{r}_0-\bm{r}'_0)\sim 1/|\bm{r}_0-\bm{r}'_0|$ 
for a large $|\bm{r}_0-\bm{r}'_0|$.
We introduce a complex fermion operator,
$\alpha(y)=\frac{1}{2}[\gamma(0,y)
+i\gamma(L,-y+y_0+y'_0)]$.
Then,
the Hamiltonian for two chiral Majorana edge states 
can be rewritten into that of the 1D chiral Dirac fermion\cite{fendley},
$\mathcal{H}_{\rm edge}=-iv\int dy \alpha^{\dagger}(y)\partial_y\alpha(y)$.
The long-range tunneling term 
is also expressed as
$\mathcal{H}_{\rm tun}=t_0(2\alpha^{\dagger}(y_0)\alpha(y_0)-1)$.
Because of the chiral character of the Dirac fermion $\alpha$, this tunneling term
raises only forward scattering, the effect of which is merely to shift the chemical potential.
As a result, the chiral Dirac fermion is still gapless.
Going back to the Majorana fields, we conclude that the two chiral Majorana edge modes are stable
against sufficiently dilute impurities.
In contrast, in the case of the $d_{xy}$-wave pairing, the long-range tunneling via nodal excitations 
exists even in the absence of impurities, as depicted in FIG.\ref{fig:edgestate}(F).
In this case,
an energy gap opens around $k_y\sim 0$, and the Majorana mode disappears 
even for a relatively large value of $L$, for which the Majorana mode still exists for the $d_{x^2-y^2}$ wave pairing.
(see FIGs.\ref{fig:edgestate}(B) and (E))

We, now, consider the stability of the Majorana mode in a vortex core.
In addition to the localized zero energy Majorana solution, 
there are also delocalized states caused by gapless nodal excitations with the finite Fermi momenta.
When there are multiple vortices in the system under consideration,
these delocalized states raise long-range tunneling between spatially separated vortices, which may
destroy the zero energy Majorana mode.
In the system with odd numbers of vortices, one Majorana mode in a vortex core survives.
However, in the case with even numbers of vortices, the Majorana mode
disappears unless they are separated enough from each other.  

{\it Topological order and topological protection of Majorana fermions ---}
The above consideration strongly implies that there is a topological order which ensures
the stability of Majorana fermion modes even for nodal superconductors with bulk gapless excitations.
However, in sharp contrast to a gapful topological order, the bulk
Chern number $\nu_{\rm Ch}$ is not well-defined for our gapless system. 
Nevertheless, we clarify here that {\it the parity of the
Chern number $(-1)^{\nu_{\rm Ch}}$ is well-defined even for
nodal superconductors}. 
The parity of the Chern number ensures the stability of the topological
order in our system.

Let us first try to define the Chern number in our gapless system. 
The simplest way to do this is to
introduce a small perturbation eliminating 
all nodes ({\it i.e.} gapless points)
in the spectrum. 
For instance, adding a small $id_{xy}$ term in the gap function, we can
easily remove all the nodes in our $d_{x^2-y^2}$ superconductor. 
After removing the nodal points, the Chern
number can be evaluated in the standard manner.
This procedure, however, does not work well after all.
The problem is that the value of the Chern number depends on the
perturbation we choose. 
As a result, one can not have a unique definition of the
Chern number for gapless systems.

On the other hand, we find that this procedure does
define the parity of the Chern number uniquely.
From the particle-hole symmetry, the parity of the Chern number is
recast into
\begin{eqnarray}
(-1)^{\nu_{\rm Ch}}={\rm exp}\left[i\int_{\Gamma_1}^{\Gamma_2}dk_i 
A_i({\bm k})+i\int_{\Gamma_3}^{\Gamma_4}dk_i 
A_i({\bm k})
\right], 
\end{eqnarray}
where $A_i({\bm k})$ is the ``gauge field'' defined by the bulk band
wave function $|u_n({\bm k})\rangle$,
$
A_i({\bm k})=\sum_n\langle u_n({\bm k})|\partial_{k_i}u_n({\bm k})\rangle, 
$
and $\Gamma_i$ is the time-reversal invariant ${\bm k}$-points,
$
\Gamma_{i=1,2,3,4}=(0,0), (\pi,0), (0,\pi), (\pi,\pi)
$ \cite{Sato10}.
Then, for the Hamiltonian (\ref{eq:hamiltonian}),
we can show that
\begin{eqnarray}
(-1)^{\nu_{\rm Ch}}
=\prod_{i=1,2,3,4}{\rm sgn}[\epsilon_{\Gamma_i}^2+\Delta_{\Gamma_i}^2-h^2], 
\label{eq:parity}
\end{eqnarray}
irrespective of the perturbation (such as $id_{xy}$ term) we choose\cite{KW09,com1}. 
This means that we have a unique value of the parity in the limit of $id_{xy}\rightarrow 0$; i.e.
the parity of the Chern number
$(-1)^{\nu_{\rm Ch}}$ is well-defined even for nodal superconductors,
although the Chern number $\nu_{\rm Ch}$ itself is not.
The parity of the Chern number 
characterizes the topological phase in nodal superconductors.
For $(-1)^{\nu_{\rm Ch}}=-1$, there exists an odd number of
topologically stable Majorana fermion in the edges and in a vortex core
for nodal superconductors. 
For example, for the model (\ref{eq:hamiltonian}) with $-4t-\mu<h<-\mu$, we obtain $(-1)^{\nu_{\rm Ch}}=-1$ from
(\ref{eq:parity}).
Thus, the existence of the gapless Majorana edge mode in
FIG.\ref{fig:edgestate} is characterized by the odd parity
$(-1)^{\nu_{\rm Ch}}=-1$.
On the other hand, for $(-1)^{\nu_{\rm Ch}}=1$, there is no
topologically stable Majorana fermion. 
We emphasize that the formula (\ref{eq:parity}) is applicable only to systems with particle-hole symmetry,
and thus, the topological order in gapless systems is specific to topological
superconducting states.

In addition to the parity of the Chern number, one can
consider another topological number dubbed 1D $Z_2$ invariant \cite{Sato10}.
The 1D $Z_2$ invariant $(-1)^{\nu[{\rm C}_{ij}]}$ is introduced as a
line integral along a specific
time-reversal invariant path ${\rm C}_{ij}$ passing through
$\Gamma_i$ and $\Gamma_j$. 
In a similar manner above, it is shown that the 1D $Z_2$ invariant is
well-defined even for our nodal superconductor, and we obtain 
$
(-1)^{\nu[{\rm C}_{ij}]}
={\rm sgn}[\epsilon_{\Gamma_i}^2+\Delta_{\Gamma_i}^2-h^2]{\rm
sgn}[\epsilon_{\Gamma_j}^2+\Delta_{\Gamma_j}^2-h^2]. 
$
For the model (\ref{eq:hamiltonian}) with $4t-\mu<h<-\mu$, this formula yields $(-1)^{\nu[{\rm C}_{12}]}=-1$ for both
$d_{x^2-y^2}$ and $d_{xy}$ superconductors.   
From the bulk-edge correspondence, 
this $Z_2$ invariant determines the location of the Majorana edge
fermions at $k_y\sim 0$, as illustrated in FIGs.\ref{fig:edgestate} (A),(B), and
(D).
Since the 1D $Z_2$ invariant is associated with a local structure in the Brillouin zone,
its non-triviality does not directly lead to the topological stability.
However, the 1D $Z_2$ invariant is
useful for identifying the location of zero energy Majorana edge modes, as shown above.
On the other hand, the odd parity of the Chern number introduced above definitely characterizes the global nontrivial topology of the Hilbert space, ensuring the topological protection mechanism of Majorana modes.

The above consideration can be straightforwardly generalized to
a general multi-band nodal superconductor.
In this case,
when the superconducting gap vanishes or becomes sufficiently small at the
time-reversal invariant ${\bm k}$-points $\Gamma_i$, the parity of the
Chern number is evaluated as
$
(-1)^{\nu_{\rm Ch}}=\prod_{\alpha,i=1,2,3,4}
{\rm sgn}[{\cal E}_{\alpha}(\Gamma_i)], 
$
where ${\cal E}_{\alpha}({\bm k})$ is the normal dispersion of the
superconductor.  The index $\alpha$ specifies an energy band including the spin degrees of freedom.
Therefore, when the Fermi level is located close to odd numbers of 
time-reversal invariant $\bm{k}$-points, and the superconducting gap vanishes
at these points because of the symmetry requirement, the NCS nodal superconductor possesses topologically protected Majorana fermion modes under an applied small magnetic field.

{\it Experimental detection of Majorana fermions ---}
For the experimental detection of Majorana fermions in nodal superconductors, one promising approach is
to exploit an interferometry measurement proposed for
a superconductor-topological-insulator junction in refs.
\cite{ANB,FK09}. 
We consider a setup similar to those proposals, but with
a difference that, instead of
a superconductor-topological-insulator junction,
a bulk $d$-wave Rashba superconductor is used.
The contribution from 
nodal excitations to the conductance in the $d$-wave pairing state vanishes
like $\sim T$ at sufficiently low temperatures,
and thus, the current is dominated by that carried by two Majorana edge modes.
The dependence of the conductance on the parity of vorticity inside the superconductor signifies clearly
the Majorana fermion contributions\cite{ANB,FK09}.

The non-Abelian nodal superconductor considered here is also
realizable in an interface between a centrosymmetric nodal superconductor such as High-$T_c$ cuprates
and a semiconductor, as considered in the case of the $s$-wave pairing state
by Sau et al. and Alicea\cite{SAU,ALICEA}. 
In such a system, because of the considerably large superconducting gap,
 the experimental detection of Majorana modes may be easier.

{\it Summary ---} 
We have demonstrated that even in nodal superconductors such as the $d$-wave pairing state,
Majorana fermion modes, 
which are topologically protected against weak perturbations and lead to the non-Abelian statistics, are realized
under a certain realistic condition, 
in spite of the existence of bulk gapless nodal excitations.
Our results establish a concept of a gapless topological phase, and open the possibility of
detecting Majorana fermions
in various NCS superconductors with gap-nodes
found in real materials.

This work is supported by the Grant-in-Aids for
Scientific Research from MEXT of Japan
(Grants No.19052003 and No.21102510 (S.F.), No.22540383 and No.22103005 (M.S.)).

\bibliography{d-wave}

\begin{thebibliography}{35}
\bibitem[{com()}]{com1}
\bibinfo{note}{More generally, we find that the right-hand side of
  (\ref{eq:parity}) is $\prod_{i}(-1)^{\zeta_{\Gamma_i}}$ with the topological
  index $\zeta_{\rm k}$ in ref.\cite{KW09}.}
\end{thebibliography}

\end{document}